\documentclass[fleqn,usenatbib]{mnras}

\usepackage{newtxtext,newtxmath}

\usepackage[T1]{fontenc}
\usepackage{ae,aecompl}

\usepackage[]{rotating}

\usepackage{graphicx}	
\usepackage{amsmath}	
\usepackage{amssymb}	

\usepackage{longtable}
\usepackage{lipsum}

\newcommand{\teff}{T$_{\rm eff}$}
\newcommand{\teq}{T$_{\rm eq}$}
\newcommand{\logg}{$\log${(g)}}
\newcommand{\met}{$[$Fe/H$]$}

\newcommand{\tduration}{$\rm{t_d}$}

\title[Synergetic photometry of HJ atmospheres]{Synergies between space telescopes in the photometric characterization of the atmospheres of Hot Jupiters}

\author[Vikash Singh et al.]{
Vikash Singh,$^{1,2}$\thanks{E-mail: vikash.singh@inaf.it}
G. Scandariato,$^{2}$
I. Pagano$^{2}$
\\
$^{1}$Dip. di Fisica e Astronomia \lq\lq Ettore Majorana\rq\rq, Universit\`a di Catania, Via S.Sofia 64, I-95123, Catania, Italy
\\
$^{2}$INAF -- Osservatorio Astrofisico di Catania, Via S.Sofia 78, I-95123, Catania, Italy\\
}

\date{Accepted 2019 May 2. Received 2019 April 18; in original form 2019 March 22}

\pubyear{2019}

\begin{document}
\label{firstpage}
\pagerange{\pageref{firstpage}--\pageref{lastpage}}
\maketitle

\begin{abstract}
Previous generation of instruments have the opportunity to discover thousands of extra-solar planets, and more will come with the current and future planet-search missions. In order to go one step further in the characterization of exoplanets, in this paper we describe a way to compare the photometric observation of Hot Jupiters done with space telescopes such as HST, CHEOPS, TESS, PLATO and JWST and give the first hand characterization on their atmospheres. We analyze a set of planetary systems hosting a Hot Jupiter for which an atmospheric template is available in literature. For each system, we simulate the transit light curves observed by different instruments, convolving the incoming spectrum with the corresponding instrumental throughput. For each instrument, we thus measure the expected transit depth and estimate the associated uncertainty. Finally, we compare the transit depths as seen by the selected instruments and we quantify the effect of the planetary atmosphere on multi-band transit photometry. We also analyze a set of simulated scenarios with different stellar magnitudes, activity levels, transit durations and atmospheric templates to find the best cases for this kind of observational approach. We find in general that current and especially future space telescopes provide enough photometric precision to detect significant differences between the transit depths at different wavelengths. In particular, we find that the chromatic effect due to the atmosphere of the Hot Jupiters is maximized at later spectral types, and that the effect of stellar activity is smaller than the measurement uncertainties.

\end{abstract}

\begin{keywords}
techniques: photometric -- stars -- planetary systems
\end{keywords}



\section{Introduction}

HD~209458b is the first exoplanet ever discovered with the transit method \citep{Charbonneau2000}. After that, the number of known transiting planets has increased especially due to dedicated space missions, such as CoRoT \citep{Baglin2006}
and Kepler (Borucki et al. 2010), and the pace is going to undergo a speed up thanks to TESS \citep{Ricker2016} and PLATO \citep{Rauer2014}. The transit technique favors large planets (whose size is similar to Jupiter) in close-in orbits, because the probability of having a planetary transit over the stellar disk increases with increasing planetary radius and decreasing orbital distance \citep{Haswell2010}. For this reason, most of the transiting planets are characterized by short semi-major axis and, by consequence, high stellar irradiation. This is why this planets are usually called as Hot Jupiters (HJs).

In the last years, these planets started to be characterized in terms of atmospheric composition. For example, some studies have been able to detect the absorption signal of atomic lines such as neutral Na and K \citep[e.g.][]{Lendl2017} or molecular bands such as water and CO \citep[e.g.][]{Brogi2016} in the atmospheres of HJs.

While spectroscopy is useful for the chemical characterization of the atmospheres, broad-band photometry is helpful in the investigation of continuum effects, e.g. Rayleigh scattering. For example, \citet{Nascimbeni2013} observed the transit of the hot-Neptune GJ4370b simultaneously in the U and R photometric bands, measuring a transit depth difference ascribed to Rayleigh scattering through a hazy atmosphere.

In this work we investigate the possibility to carry on an observational program aimed at gathering simultaneous transit light curves using the latest space telescopes. As a matter of fact, with the new class of space telescopes and their improved photometric capabilities, it will be possible to characterize the atmospheres of extrasolar planets with better precision using broad band photometry. Contamination by the Earth's atmospheres will be much reduced, for the benefit of the accuracy of the measurements which, moreover, can be easily extended to the infrared part of the spectrum. Finally, this kind of studies can be extended to fainter and dimmer stars, and to smaller exoplanets, thus increasing the size of the sample of exoplanets with atmospheric characterization.

\begin{figure*}
\centering
\includegraphics[width=\linewidth]{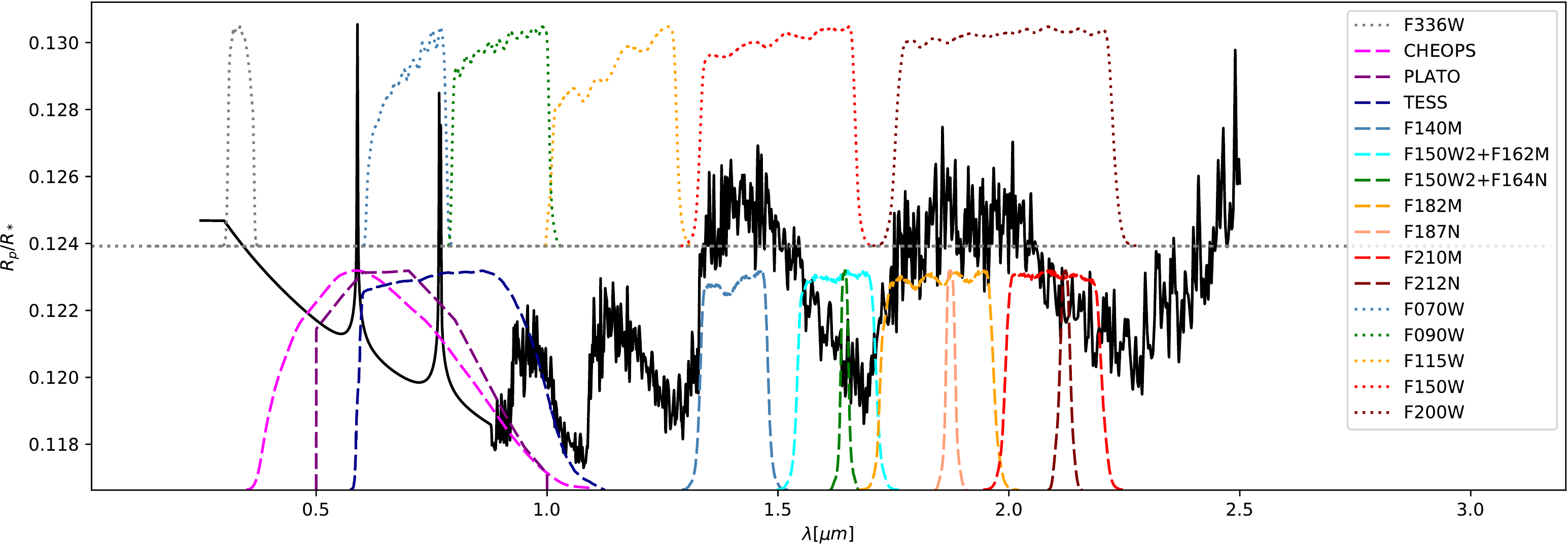}
\caption{Transmission spectrum of the HJ WASP-17b used in this work together with the throughput of the instruments in Table~\ref{tab:instruments}. WFC3's F336W and JWST's wide passbands (the ones with a trailing \lq\lq W\rq\rq\ in the filter name) have been offset up for clarity. Water bands between $\sim$1\micron\ and $\sim$2\micron\ are apparent.}\label{fig:spectrum}
\end{figure*}

In this paper we analyze the synergies between different present and future space telescopes. In particular, we focus on the possibility to detect differences in the photometric transit depth when the transits are observed using different passbands. The telescopes we take into account are the CHaracterising ExOPlanet Satellite \citep[CHEOPS,][]{Fortier2014}, the Transiting Exoplanet Survey Satellite (TESS) and the PLAnetary Transits and Oscillation of stars telescope (PLATO), which are specifically designed for the detection and characterization of exoplanets using the transit technique. We also include the James Webb Space Telescope \citep[JWST,][]{Gardner2006}, whose Near-Infrared Camera \citep[NIRCAM,][]{Horner2003} offers the possibility to perform photometric observations in the RIJH spectral interval. The full list of telescopes and passbands is reported is Table~\ref{tab:instruments}, the corresponding passbands are shown in Fig.~\ref{fig:spectrum} together with the transmission spectrum of the HJ WASP-17b \citep{Sing2016}.

\begin{table}
 \begin{center}
\caption{Instruments (and filters for WFC3@HST and NIRCAM@JWST) with corresponding effective wavelengths $\lambda_{\rm eff}$ and full width at half maximum FWHM.}\label{tab:instruments}
  \begin{tabular}{cccc}
 \hline\hline
Instrument & Filter & $\lambda_{\rm eff}$ & FWHM\\
       & & (\micron)  & (\micron)\\

 \hline
WFC3@HST & F336W & 0.336  &  0.055 \\
CHEOPS & --- &  0.646  &  0.422\\
TESS & --- &   0.798  &  0.400\\
\hline
PLATO & --- & 0.700  &  0.369\\
NIRCAM@JWST & F070W  &  0.706  &  0.160 \\
NIRCAM@JWST & F090W  &  0.904  &  0.210 \\
NIRCAM@JWST & F115W  &  1.157  &  0.268 \\
NIRCAM@JWST & F140M  &  1.406  &  0.148 \\
NIRCAM@JWST & F150W  &  1.504  &  0.337 \\
NIRCAM@JWST & F150W2+F162M  &  1.628  &  0.171 \\
NIRCAM@JWST & F150W2+F164N  &  1.645  &  0.018 \\
NIRCAM@JWST & F182M  &  1.847  &  0.246 \\
NIRCAM@JWST & F187N  &  1.874  &  0.021 \\
NIRCAM@JWST & F200W  &  1.993  &  0.472 \\
NIRCAM@JWST & F210M  &  2.096  &  0.209 \\
NIRCAM@JWST & F212N  &  2.121  &  0.025 \\
\hline
  \end{tabular}
 \end{center}
\end{table}
\section{Computation of synthetic photometry}\label{sec:photometry}

For our analysis we use the spectral library of the \lq\lq Comparing the Range of Exoplanet Atmospheres with Transmission and Emission Spectra\rq\rq\ (CREATES) project\footnote{\url{https://pages.jh.edu/~dsing3/David_Sing/Spectral_Library.html}} as a representative collection of the different atmospheres of the HJs known so far. The list of HJs and the corresponding references for the spectral models are reported in Table~\ref{tab:hjs}. These spectral models are given in units of $k(\lambda)=(R_p/R_*)_\lambda$ ($R_p$ is the planetary radius, $R_*$ is the stellar radius) as a function of wavelength. The second power of these spectra (i.e.\ $k(\lambda)^2$) is the fraction of stellar light blocked by the planet during the transit as a function of wavelength, i.e.\ the transit spectrum.

\begin{table*}
 \begin{center}
\caption{HJs in the CREATES database and analyzed in this work. The $\Delta Z_{UB-LM}$ index is increases with the strength of Rayleigh scattering in the atmosphere \citep{Sing2016}.}\label{tab:hjs}
  \begin{tabular}{lcl}
 \hline\hline
Planet & $\Delta Z_{UB-LM}$ & ref.\\
 \hline
WASP-17b  &  -0.8  & \citet{Fortney2010,Sing2016}\\
WASP-31b  &  2.15  & \citet{Fortney2010,Sing2013,Sing2016} \\
HD209458b  &  0.73  & \citet{Fortney2010,Sing2016}\\
HAT-P-1b  &  2.01  & \citet{Fortney2010,Nikolov2014,Sing2016}\\
WASP-19b  &  1.04  & \citet{Fortney2010,Sing2016}\\
WASP-12b  &  3.76  & \citet{Sing2013,Sing2016}\\
HAT-P-12b  &  4.14  & \citet{Fortney2010,Sing2016}\\
HD189733b  &  5.52  & \citet{Fortney2010,Sing2016}\\
WASP-6b  &  8.49  & \citet{Fortney2010,Nikolov2015,Sing2016}\\
WASP-39b  &  0.1  & \citet{Fortney2010,Sing2016}\\
WASP-121b  &  ---  & \citet{Evans2016}\\
HAT-P-26b  &  ---  & \citet{Wakeford2017}\\
\hline
  \end{tabular}
 \end{center}
\end{table*}

For each HJs, we model the out-of-transit spectrum $f(\lambda)$ of the corresponding host star by using the \teff\ reported in literature (see Table~\ref{tab:hjs}) and selecting the closest match in the BT-Settl synthetic spectral library \citep{Baraffe2015}, assuming \logg=5.0, \met=0 and no $\alpha$ enhancement.

From the observer's point of view, the flux observed during the transit $f^\prime(\lambda)$ is the difference between the stellar spectrum and the spectrum of the light blocked by the planet, i.e.:
\begin{equation}
f^\prime(\lambda)=f(\lambda)-f(\lambda)k(\lambda)^2
\end{equation}

To estimate the transit depth $\delta$ in relative units of flux, we convolve $f(\lambda)$ and $f^\prime(\lambda)$ with the instrumental throughput $\Gamma(\lambda)$ (which includes both optical throughput and quantum efficiency) of the facilities listed in Table~\ref{tab:instruments}, obtaining:
\begin{equation}
\delta=\frac{\int_\lambda \Gamma(\lambda)\left(f(\lambda)-f^\prime(\lambda)\right)}{\int_\lambda \Gamma(\lambda)f(\lambda)}=\frac{\int_\lambda \Gamma(\lambda)f(\lambda)k(\lambda)^2}{\int_\lambda \Gamma(\lambda)f(\lambda)}
\end{equation}

We remark that we intentionally neglect any nightside pollution of the in-transit photometry. This is supported by \citet{Kipping2010}, which estimate that the contribution of the thermal emission from the planetary nightside is less than 10 ppm for a typical HJ at wavelength shorter that 2\micron. As we will find below, this is negligible compared with the typical transit depth uncertainty.

For the synthetic photometry of CHEOPS, we obtain the photometric uncertainty on a 1~hr observation by using the CHEOPSim simulator\footnote{\url{https://cheops.unige.ch/cheopsim/}}. For TESS, we compute the 1-hr uncertainty using the \texttt{ticgen} package for Python \citep{Jaffe2017}. For PLATO we use the 1-hr noise vs.\ magnitude relation provided by \citet{Rauer2014} in the conservative case that the star is observed with the minimum number (i.e., 8) of cameras simultaneously. For the passbands of NIRCAM@JWST, we use the Exposure Time Calculator (ETC) \lq\lq Pandeia Engine\rq\rq\footnote{\url{https://jwst-docs.stsci.edu/display/JPP/JWST+ETC+Pandeia+Engine+Tutorial}} version 1.3 configuring NIRCAM in \lq\lq short-wavelength\rq\rq\ mode, with the \lq\lq rapid\rq\rq\ read-out mode of the smallest subarray allowed by the electronics (\lq\lq sub64p\rq\rq). These settings allow the observation of bright stars avoiding saturation \citep{Robberto2010,Stansberry2014}. We run the ETC iteratively to find the maximum exposure time which maximizes the S/N of the photometry without reaching saturation. Once the best exposure time is found, the photometric uncertainty is rescaled to a nominal 1-hr observation.

We compute the transit depth $\delta$ as the difference between the in-transit and out-of-transit stellar photometry (Table~\ref{tab:depths}). The corresponding uncertainty is obtained by rescaling the 1-hr photometric uncertainty with the square root of the transit time duration \tduration\ (we assume poissonian noise). In the case of CHEOPS and JWST, the out-of-transit photometry will cover a timespan similar to the transit duration: for this reason, we compute the uncertainty on the transit depth by multiplying the photometric uncertainty by $\sqrt{2}$. In the case of PLATO and TESS, the photometric time series will run much longer than the orbital period of the HJs: we thus assume that the uncertainty on the out-of-transit photometry is negligible with respect to the in-transit photometry, and we do not apply any correction to the uncertainty on the transit depth.

For the brightest stars, we find that saturation cannot be avoided for JWST's observations, in particular in the wide and medium bands. These measurements are flagged with \lq\lq\texttt{nan}\rq\rq\ uncertainties in Table~\ref{tab:depths}.

\begin{table*}
 \begin{center}
\caption{Simulated transit depths as seen by the instruments in Table~\ref{tab:instruments} for the HJs in the CREATES database. Measurements with \texttt{nan} uncertainties are the ones for which saturation cannot be avoided.}\label{tab:depths}
  \begin{tabular}{lcccccccc}
 \hline\hline
Planet & $\delta_{F336W}$ & $\delta_{CHEOPS}$ & $\delta_{PLATO}$ & $\delta_{TESS}$ & $\delta_{F140M}$ & $\delta_{F150W2+F162M}$ & $\delta_{F150W2+F164N}$ & $\delta_{F182M}$ \\
       & (ppm)  & (ppm)  & (ppm) & (ppm)& (ppm) & (ppm) & (ppm) & (ppm) \\

 \hline
 \\
WASP-17b & 15380 $\pm$ 59 & 14588 $\pm$ 105 & 14503 $\pm$ 38 & 14384 $\pm$ 163 & 15496 $\pm$ 22 & 14698 $\pm$ 23 & 14506 $\pm$ 66 & 15389 $\pm$ 23
 \\
WASP-31b & 16555 $\pm$ 83 & 15725 $\pm$ 147 & 15697 $\pm$ 52 & 15687 $\pm$ 247 & 15812 $\pm$ 33 & 15643 $\pm$ 35 & 15636 $\pm$ 102 & 15804 $\pm$ 34 
 \\
HD209458b & 14955 $\pm \mathrm{nan}$  & 14760 $\pm$ 34 & 14746 $\pm$  7 & 14700 $\pm$ 43 & 14775 $\pm \mathrm{nan}$  & 14635 $\pm \mathrm{nan}$  & 14633 $\pm \mathrm{nan}$  & 14724 $\pm \mathrm{nan}$ 
 \\
HAT-P-1b & 14066 $\pm$ 43 & 13855 $\pm$ 59 & 13838 $\pm$ 26 & 13802 $\pm$ 107 & 14017 $\pm \mathrm{nan}$  & 13756 $\pm \mathrm{nan}$  & 13703 $\pm$ 45 & 13971 $\pm \mathrm{nan}$ 
 \\
WASP-19b & 19805 $\pm$ 198 & 19528 $\pm$ 238 & 19510 $\pm$ 108 & 19482 $\pm$ 394 & 19891 $\pm$ 42 & 19651 $\pm$ 43 & 19598 $\pm$ 122 & 19877 $\pm$ 40 
 \\
WASP-12b & 14600 $\pm$ 77 & 14388 $\pm$ 139 & 14364 $\pm$ 49 & 14329 $\pm$ 202 & 14205 $\pm$ 27 & 14164 $\pm$ 28 & 14161 $\pm$ 80 & 14128 $\pm$ 27  
 \\
HAT-P-12b & 20571 $\pm$ 329 & 19590 $\pm$ 195 & 19546 $\pm$ 102 & 19381 $\pm$ 339 & 18939 $\pm$ 31 & 18408 $\pm$ 31 & 18374 $\pm$ 88 & 18630 $\pm$ 29 
 \\
HD189733b & 24906 $\pm \mathrm{nan}$  & 24366 $\pm$ 44 & 24328 $\pm$  9 & 24249 $\pm$ 53 & 24060 $\pm \mathrm{nan}$  & 23780 $\pm \mathrm{nan}$  & 23760 $\pm \mathrm{nan}$  & 23897 $\pm \mathrm{nan}$   
 \\
WASP-6b & 21593 $\pm$ 113 & 20971 $\pm$ 174 & 20919 $\pm$ 58 & 20824 $\pm$ 255 & 20655 $\pm$ 31 & 20306 $\pm$ 31 & 20275 $\pm$ 89 & 20488 $\pm$ 30
 \\
WASP-39b & 21753 $\pm$ 120 & 20909 $\pm$ 178 & 20856 $\pm$ 62 & 20719 $\pm$ 246 & 21565 $\pm$ 28 & 20547 $\pm$ 29 & 20344 $\pm$ 82 & 21384 $\pm$ 27  
 \\
WASP-121b & 14323 $\pm$ 42 & 15391 $\pm$ 60 & 15398 $\pm$ 26 & 15256 $\pm$ 119 & 14772 $\pm$ 22 & 14427 $\pm$ 23 & 14353 $\pm$ 54 & 14665 $\pm$ 21
 \\
HAT-P-26b & 4995 $\pm$ 123 & 4938 $\pm$ 153 & 4937 $\pm$ 54 & 4946 $\pm$ 200 & 5228 $\pm$ 26 & 4960 $\pm$ 25 & 4947 $\pm$ 66 & 5144 $\pm$ 24

 \\

\hline
  \end{tabular}
  
\begin{tabular}{lccccccccc}
 \hline\hline
Planet & ... & $\delta_{F187N}$ & $\delta_{F210M}$ & $\delta_{F212N}$ & $\delta_{F070W}$ & $\delta_{F090W}$ & $\delta_{F115W}$ & $\delta_{F150W}$ & $\delta_{F200W}$\\
      & &(ppm) & (ppm) & (ppm) & (ppm) & (ppm) & (ppm) & (ppm) & (ppm)\\

 \hline
 \\
WASP-17b & &15655 $\pm$ 73 & 15140 $\pm$ 27 & 14905 $\pm$ 75 & 14590 $\pm$ 20 & 14287 $\pm$ 19 & 14379 $\pm$ 19 & 15253 $\pm$ 18 & 15310 $\pm$ 18
 \\
WASP-31b & &15914 $\pm$ 111 & 15712 $\pm$ 42 & 15650 $\pm$ 115 & 15844 $\pm$ 30 & 15625 $\pm$ 26 & 15637 $\pm$ 26 & 15744 $\pm$ 25 & 15774 $\pm$ 27
 \\
HD209458b & &14778 $\pm$ 19 & 14669 $\pm \mathrm{nan}$  & 14632 $\pm$ 18 & 14783 $\pm \mathrm{nan}$  & 14643 $\pm \mathrm{nan}$  & 14641 $\pm \mathrm{nan}$  & 14716 $\pm \mathrm{nan}$  & 14706 $\pm \mathrm{nan}$ 
 \\
HAT-P-1b & &14046 $\pm$ 48 & 13864 $\pm$ 20 & 13772 $\pm$ 50 & 13868 $\pm \mathrm{nan}$  & 13766 $\pm \mathrm{nan}$  & 13729 $\pm \mathrm{nan}$  & 13927 $\pm \mathrm{nan}$  & 13935 $\pm \mathrm{nan}$ 
 \\
WASP-19b & &19968 $\pm$ 132 & 19816 $\pm$ 50 & 19754 $\pm$ 138 & 19543 $\pm$ 43 & 19457 $\pm$ 35 & 19512 $\pm$ 34 & 19818 $\pm$ 32 & 19861 $\pm$ 32
 \\
WASP-12b & &14122 $\pm$ 87 & 14093 $\pm$ 33 & 14091 $\pm$ 90 & 14363 $\pm$ 25 & 14301 $\pm$ 23 & 14246 $\pm$ 23 & 14190 $\pm$ 23 & 14110 $\pm$ 22
 \\
HAT-P-12b & &18791 $\pm$ 92 & 18296 $\pm$ 36 & 18111 $\pm$ 98 & 19747 $\pm$ 38 & 19193 $\pm$ 29 & 18844 $\pm$ 27 & 18711 $\pm$ 26 & 18503 $\pm$ 24
 \\
HD189733b & &23980 $\pm \mathrm{nan}$  & 23720 $\pm \mathrm{nan}$  & 23620 $\pm \mathrm{nan}$  & 24347 $\pm \mathrm{nan}$  & 24184 $\pm \mathrm{nan}$  & 24011 $\pm \mathrm{nan}$  & 23943 $\pm \mathrm{nan}$  & 23830 $\pm \mathrm{nan}$ 
 \\
WASP-6b & &20593 $\pm$ 96 & 20278 $\pm$ 37 & 20147 $\pm$ 100 & 20933 $\pm$ 32 & 20746 $\pm$ 26 & 20553 $\pm$ 25 & 20519 $\pm$ 24 & 20406 $\pm$ 24
 \\
WASP-39b & &21672 $\pm$ 88 & 20965 $\pm$ 34 & 20612 $\pm$ 92 & 20987 $\pm$ 29 & 20586 $\pm$ 24 & 20444 $\pm$ 25 & 21203 $\pm$ 23 & 21243 $\pm$ 22
 \\
WASP-121b & &14771 $\pm$ 59 & 14597 $\pm$ 23 & 14532 $\pm$ 61 & 15472 $\pm$ 20 & 15078 $\pm \mathrm{nan}$  & 14716 $\pm \mathrm{nan}$  & 14661 $\pm \mathrm{nan}$  & 14647 $\pm$ 18
 \\
HAT-P-26b & &5218 $\pm$ 71 & 5015 $\pm$ 28 & 4961 $\pm$ 74 & 4930 $\pm$ 26 & 4961 $\pm$ 24 & 4971 $\pm \mathrm{nan}$  & 5115 $\pm \mathrm{nan}$  & 5099 $\pm$ 21

 \\

\hline
  \end{tabular}  
  
 \end{center}
\end{table*}

\section{Analysis of known systems}\label{sec:known}

We compare the transit depths at different wavelengths looking for differences which may indicate the presence of a non-gray atmosphere around the HJs. Since non-simultaneous observation of transits may be affected by stellar activity, the best way to follow this approach is to observe the same transits with different instruments.

The first couple of instruments we test is made up by CHEOPS and TESS, as they will operate simultaneously for a few years in the near future (TESS is already flying, while CHEOPS will be launched in the last quarter of 2019). While TESS is scanning almost the whole sky following a fixed schedule, CHEOPS will be able to point at targets on demand. This leads to the possibility of scheduling the observation of a given HJ with CHEOPS when it is already in TESS' field of view. The original idea to compare CHEOPS' and TESS' observations is already discussed in \citet{Gaidos2017}, but their work lacks considerations on the photometric precision of the observations.

\begin{figure*}
\centering
\includegraphics[width=\linewidth]{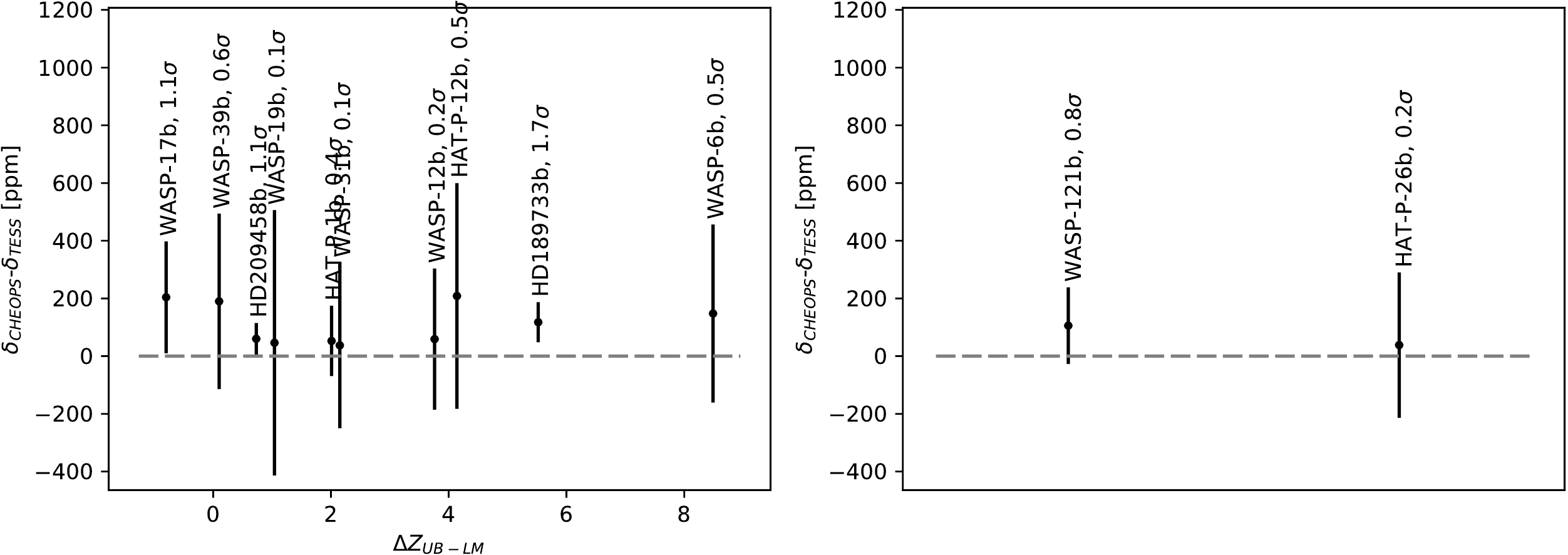}
\caption{\textit{Left panel - }Transit depth difference $\delta_{\rm CHEOPS}-\delta_{\rm TESS}$ as a function of the blue-optical to mid-IR altitude difference $\Delta Z_{\rm UB-LM}$. Labels indicate the name of the HJ and the significance level of the detection. \textit{Right panel - }Transit depth difference for the HJ with unknown $\Delta Z_{\rm UB-LM}$.}\label{fig:cheops_tess}
\end{figure*}

The CHEOPS vs.\ TESS comparison is shown in Fig.~\ref{fig:cheops_tess}. In particular, we plot the difference between the transit depths $\delta_{\rm CHEOPS}-\delta_{\rm TESS}$ as a function of the blue-optical to mid-IR altitude difference $\Delta Z_{\rm UB-LM}$ as defined by \citet{Sing2016}, where the latter index increases with the strength of Rayleigh scattering in the atmosphere (i.e.\ $\Delta Z_{\rm UB-LM}$ increases with the presence of scattering hazes and decreases with the strength of H$_2$O, CO and CH$_4$ absorption bands in the mid-IR). As $\Delta Z_{\rm UB-LM}$ is not reported for all the HJs in the CREATES project, we plot the remaining HJs with unknown $\Delta Z_{\rm UB-LM}$ in the right panel of Fig.~\ref{fig:cheops_tess}.

\begin{figure*}
\centering
\includegraphics[width=\linewidth]{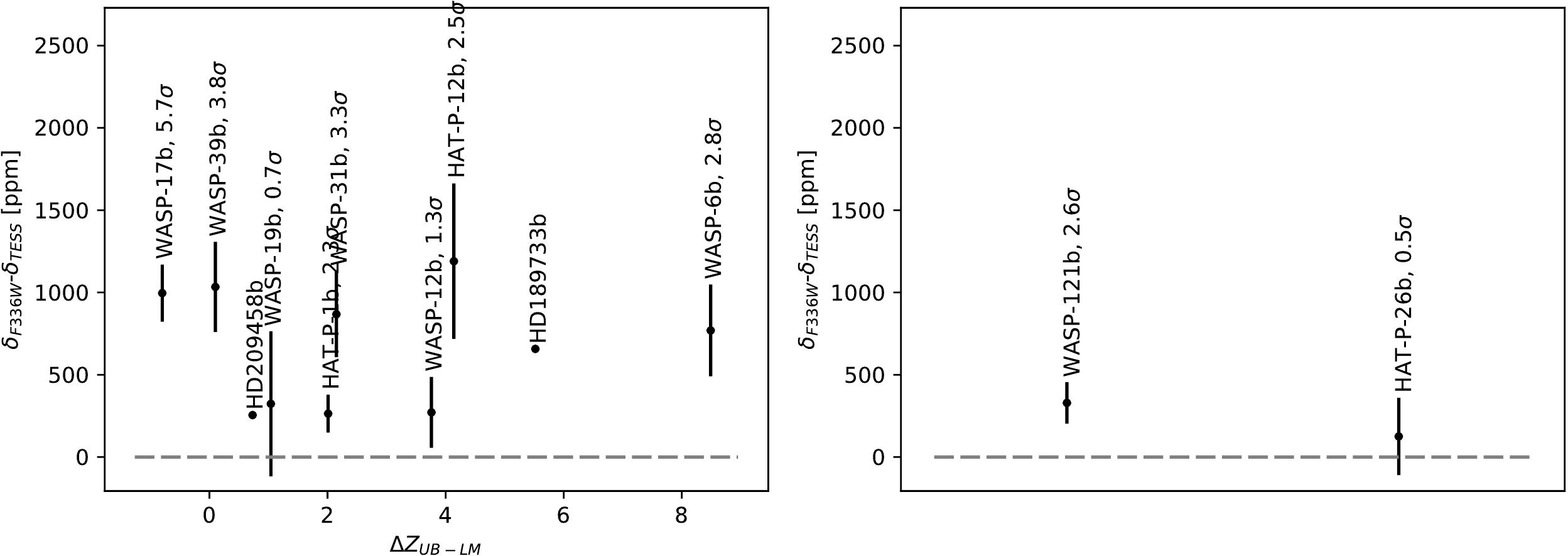}
\caption{Same as in Fig.~\ref{fig:cheops_tess} for the $\delta_{F336W}-\delta_{TESS}$ transit depth difference. Data points with no error bars are saturated even in the shortest available exposure for WFC3.}\label{fig:f336w_tess}
\end{figure*}

We generally find that the difference between the transit depths is between 0 and 300~ppm, with a significance of the detection below 2$\sigma$. This is due to the fact that the wavelength leverage between the two telescopes is too short, such that the two passbands largely overlap. This makes it difficult to detect the decrease of the transit depth with increasing wavelength typical of scattering atmospheres, especially at the V magnitudes of the stars in the CREATES database (typically V$>$10). As we discuss in Sect.~\ref{sec:activity}, significant differences in transit depths between CHEOPS and TESS can be measured for cool bright stars.

Another possibility is to increase the wavelength distance by using the UV capabilities of the Wide Field Camera 3 \citep[WFC3, ][]{Leckrone1998} of the Hubble Space Telescope (HST). To this purpose, we select the bluest filter of WFC3 whose spectral range is covered by the atmospheric model of the HJs under study, namely F336W. In Fig.~\ref{fig:f336w_tess} we plot the transit depth difference as measured by WFC3 and TESS: we find that increases by a factor of $\sim$4-5, such that in the most favorable cases is significant at the 3$\sigma$ level. Replacing TESS with CHEOPS we obtain similar results, due to the fact that the two passbands overlap with each other. Another option is to increase the wavelength distance using the IR passbands of WFC3. We do not investigate this possibility here, as the comparison between optical and IR observations is discussed extensively below for the PLATO vs.\ JWST case, whose photometric precision will allow more robust results.

To enlarge the wavelength leverage, we now compare the simulation done for PLATO and JWST, as these telescopes will likely operate simultaneously in the next decade. Also in this case, it is plausible to schedule JWST's observations in the same direction of PLATO, the latter having a fixed pattern of sky fields to follow.

\begin{figure*}
\centering
\includegraphics[width=\linewidth]{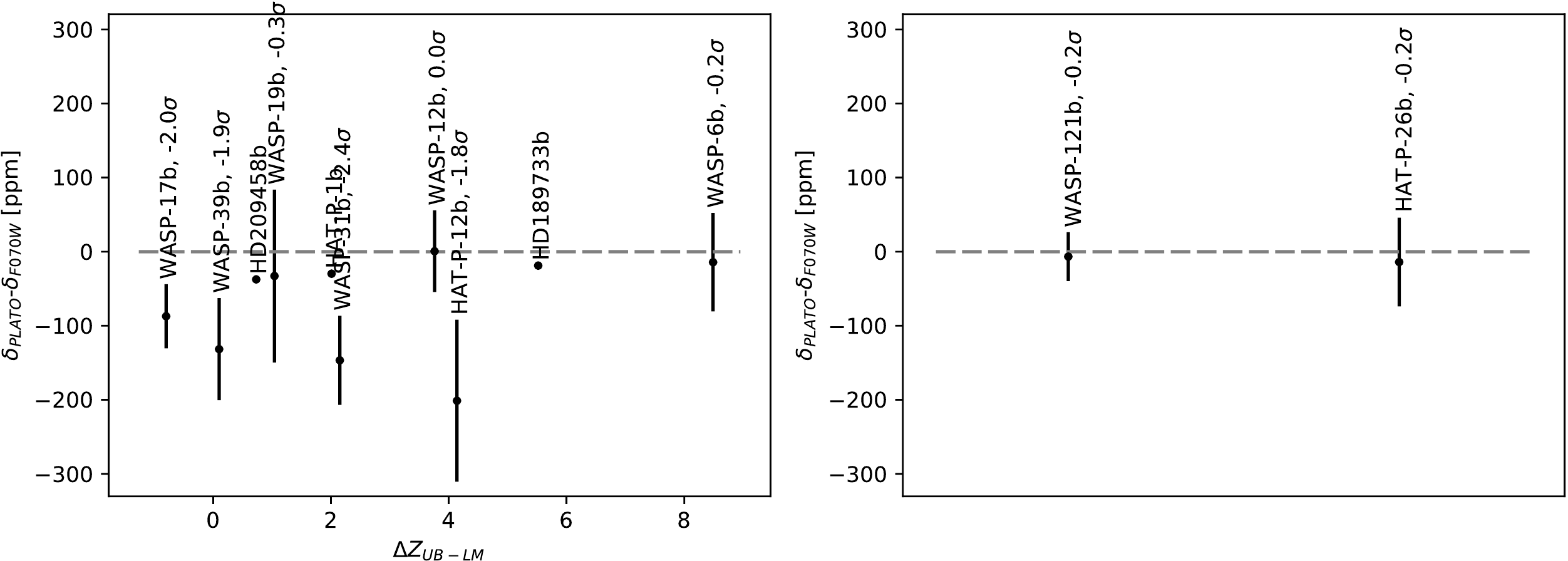}
\caption{Same as in Fig.~\ref{fig:cheops_tess} for the $\delta_{\rm PLATO}-\delta_{\rm F070W}$ transit depth difference. Data points with no error bars are saturated even in the shortest available exposure for NIRCAM.}\label{fig:plato_jw070}
\end{figure*}

\begin{figure*}
\centering
\includegraphics[width=\linewidth]{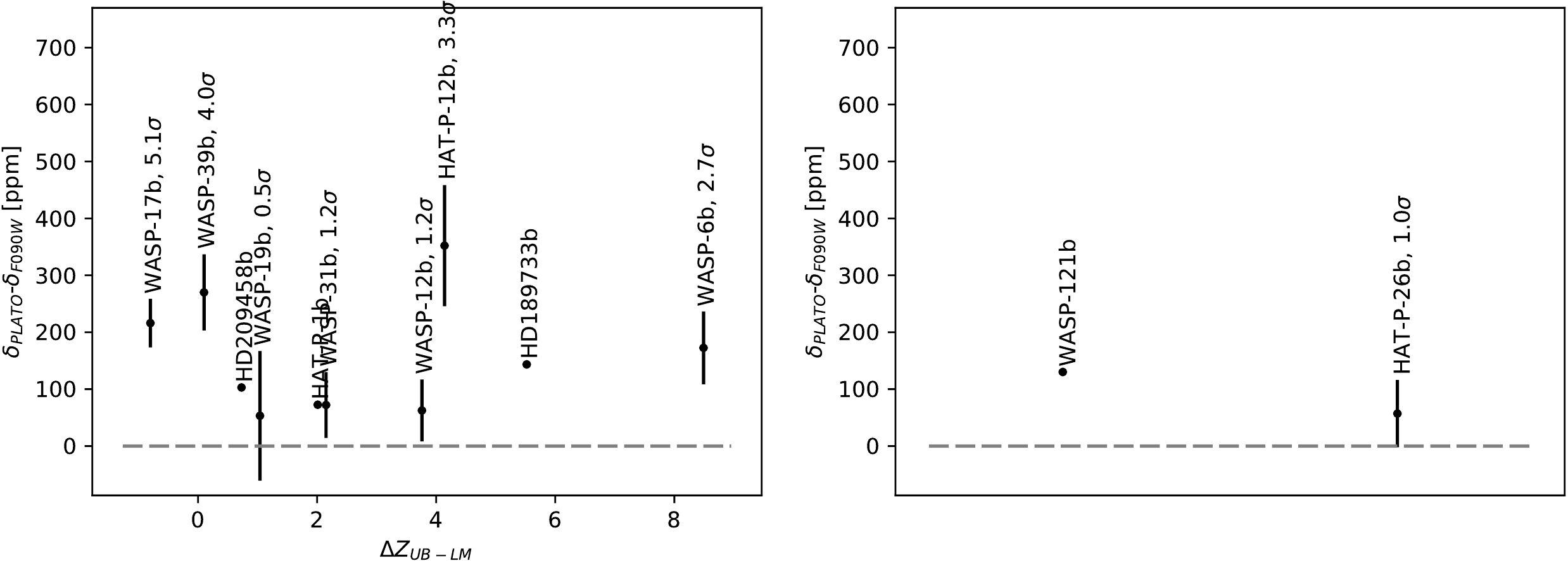}
\caption{Same as in Fig.~\ref{fig:cheops_tess} for the $\delta_{\rm PLATO}-\delta_{\rm F090W}$ transit depth difference. Data points with no error bars are saturated even in the shortest available exposure for NIRCAM.}\label{fig:plato_jw090}
\end{figure*}

\begin{figure*}
\centering
\includegraphics[width=\linewidth]{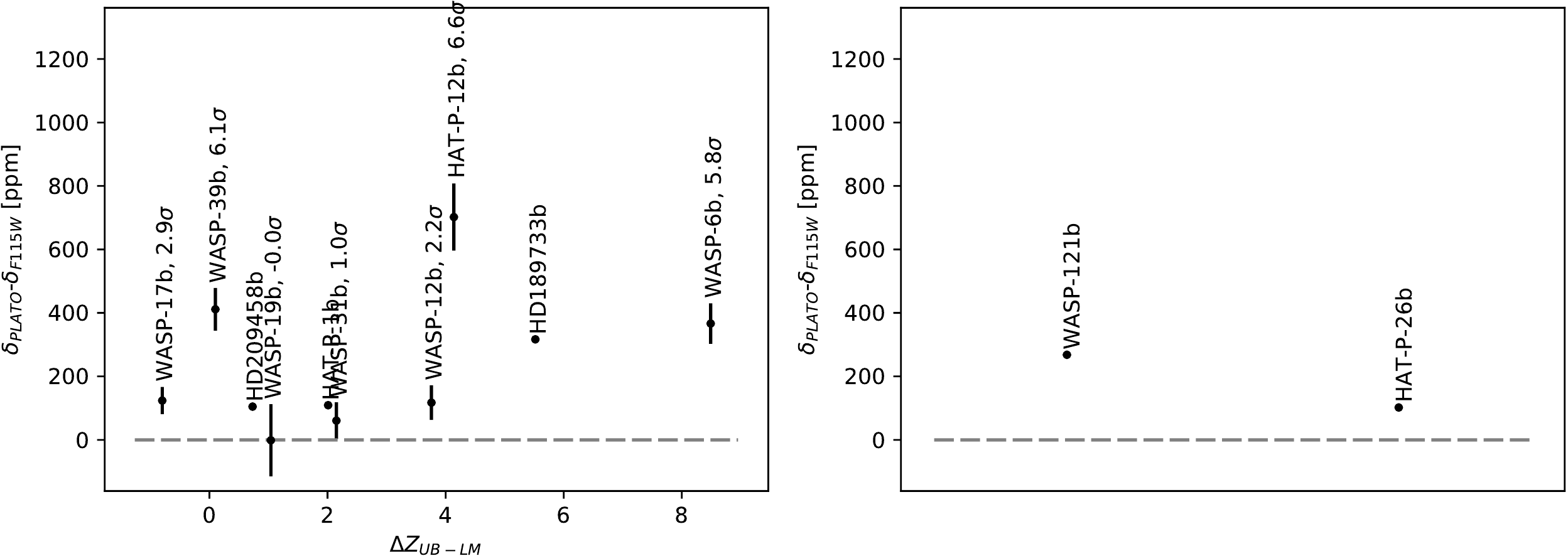}
\caption{Same as in Fig.~\ref{fig:cheops_tess} for the $\delta_{\rm PLATO}-\delta_{\rm F115W}$ transit depth difference. Data points with no error bars are saturated even in the shortest available exposure for NIRCAM.}\label{fig:plato_jw115}
\end{figure*}

\begin{figure*}
\centering
\includegraphics[width=\linewidth]{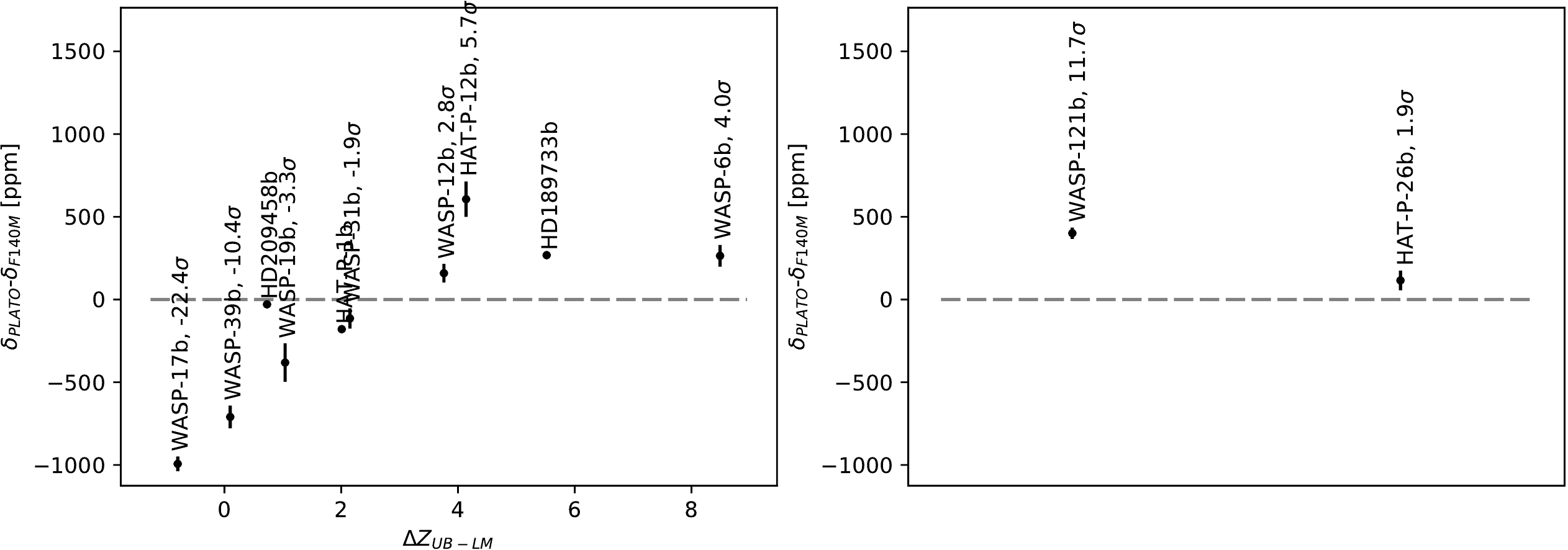}
\caption{Same as in Fig.~\ref{fig:cheops_tess} for the $\delta_{\rm PLATO}-\delta_{\rm F140M}$ transit depth difference. Data points with no error bars are saturated even in the shortest available exposure for NIRCAM.}\label{fig:plato_jw140}
\end{figure*}

\begin{figure*}
\centering
\includegraphics[width=\linewidth]{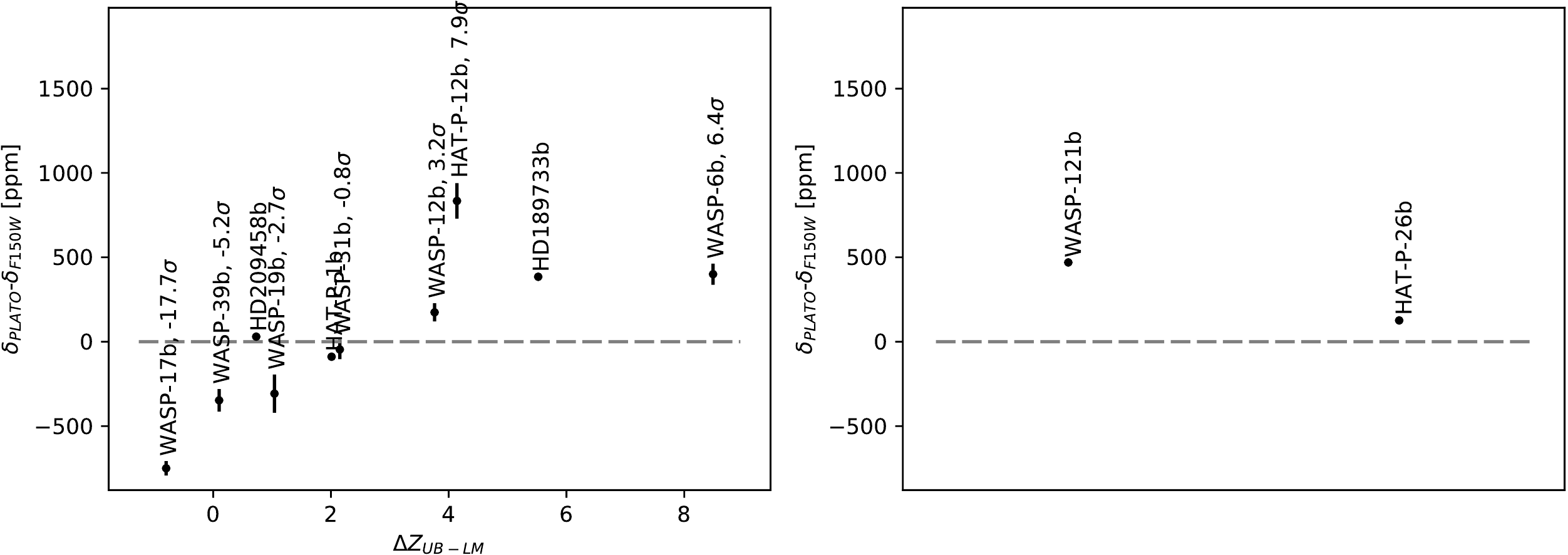}
\caption{Same as in Fig.~\ref{fig:cheops_tess} for the $\delta_{\rm PLATO}-\delta_{\rm F150W}$ transit depth difference. Data points with no error bars are saturated even in the shortest available exposure for NIRCAM.}\label{fig:plato_jw150}
\end{figure*}

\begin{figure*}
\centering
\includegraphics[width=\linewidth]{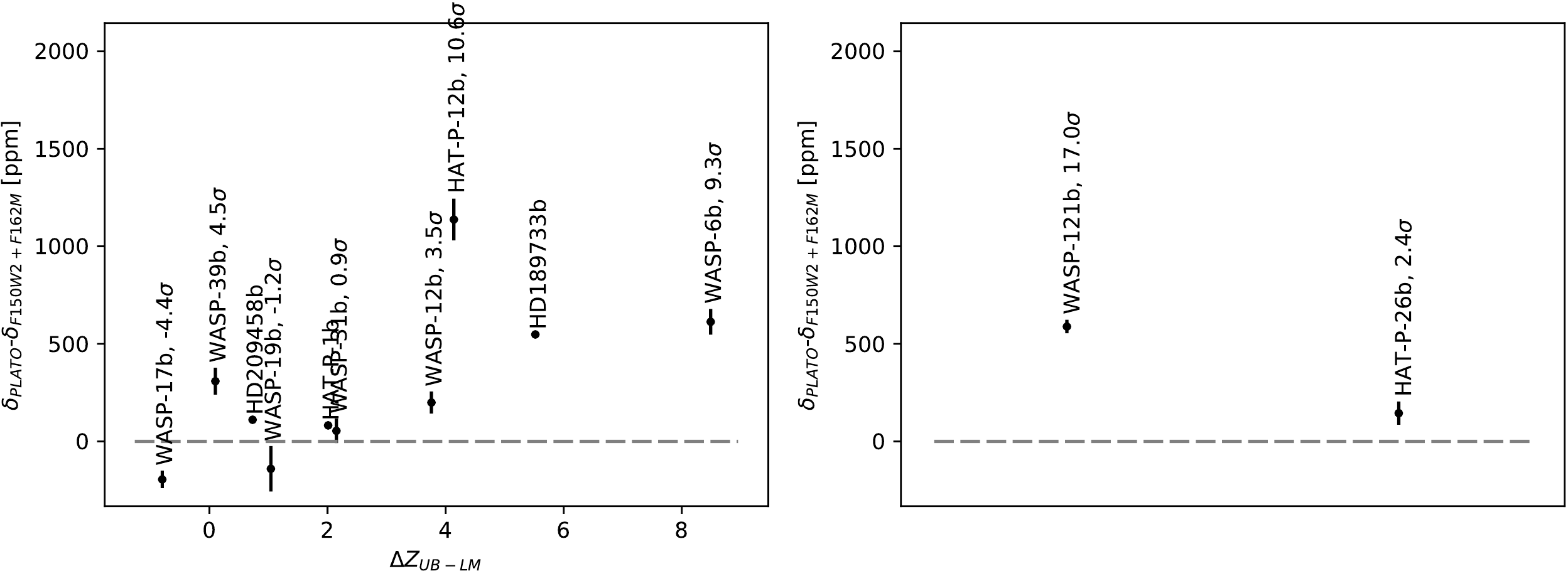}
\caption{Same as in Fig.~\ref{fig:cheops_tess} for the $\delta_{\rm PLATO}-\delta_{\rm F150W2+F162M}$ transit depth difference. Data points with no error bars are saturated even in the shortest available exposure for NIRCAM.}\label{fig:plato_jw162}
\end{figure*}

\begin{figure*}
\centering
\includegraphics[width=\linewidth]{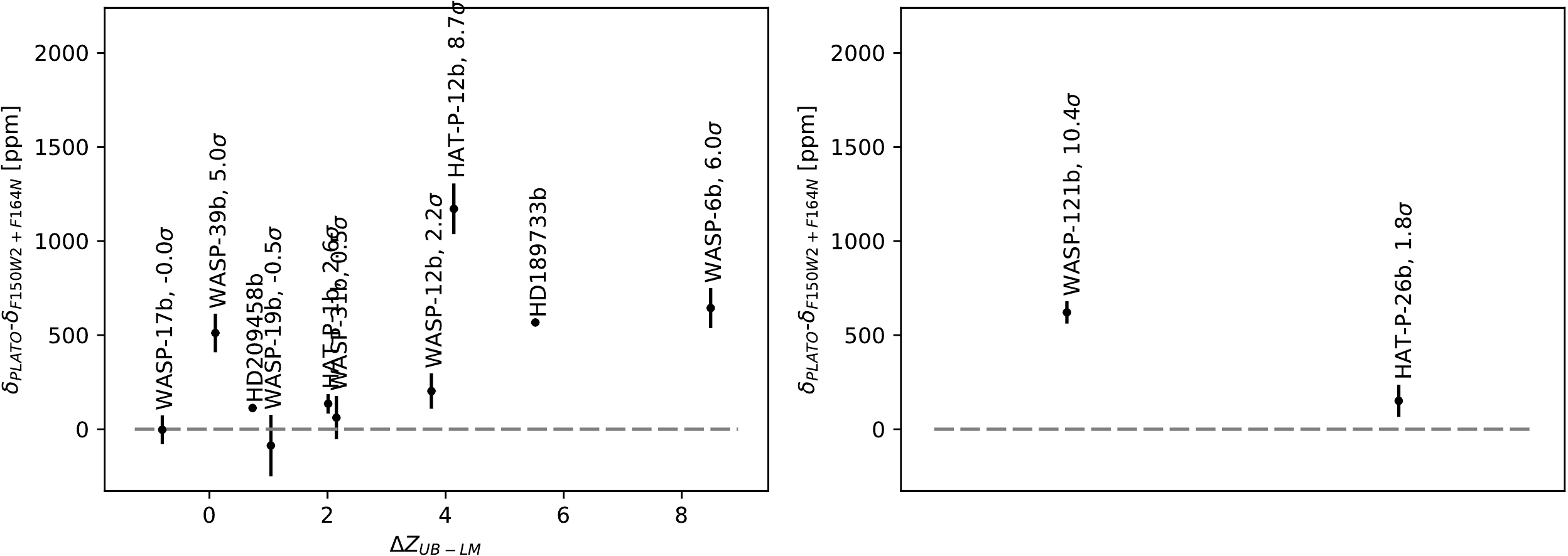}
\caption{Same as in Fig.~\ref{fig:cheops_tess} for the $\delta_{\rm PLATO}-\delta_{\rm F150W2+F164N}$ transit depth difference. Data points with no error bars are saturated even in the shortest available exposure for NIRCAM.}\label{fig:plato_jw164}
\end{figure*}

\begin{figure*}
\centering
\includegraphics[width=\linewidth]{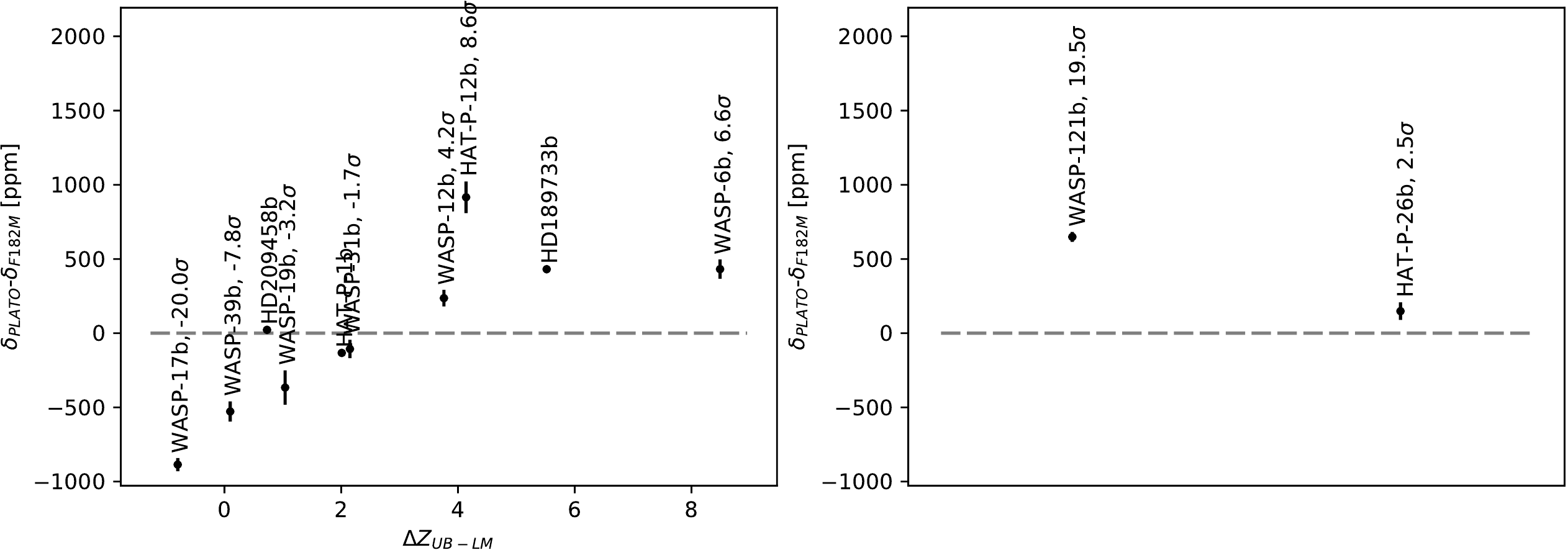}
\caption{Same as in Fig.~\ref{fig:cheops_tess} for the $\delta_{\rm PLATO}-\delta_{\rm F182M}$ transit depth difference. Data points with no error bars are saturated even in the shortest available exposure for NIRCAM.}\label{fig:plato_jw182}
\end{figure*}

\begin{figure*}
\centering
\includegraphics[width=\linewidth]{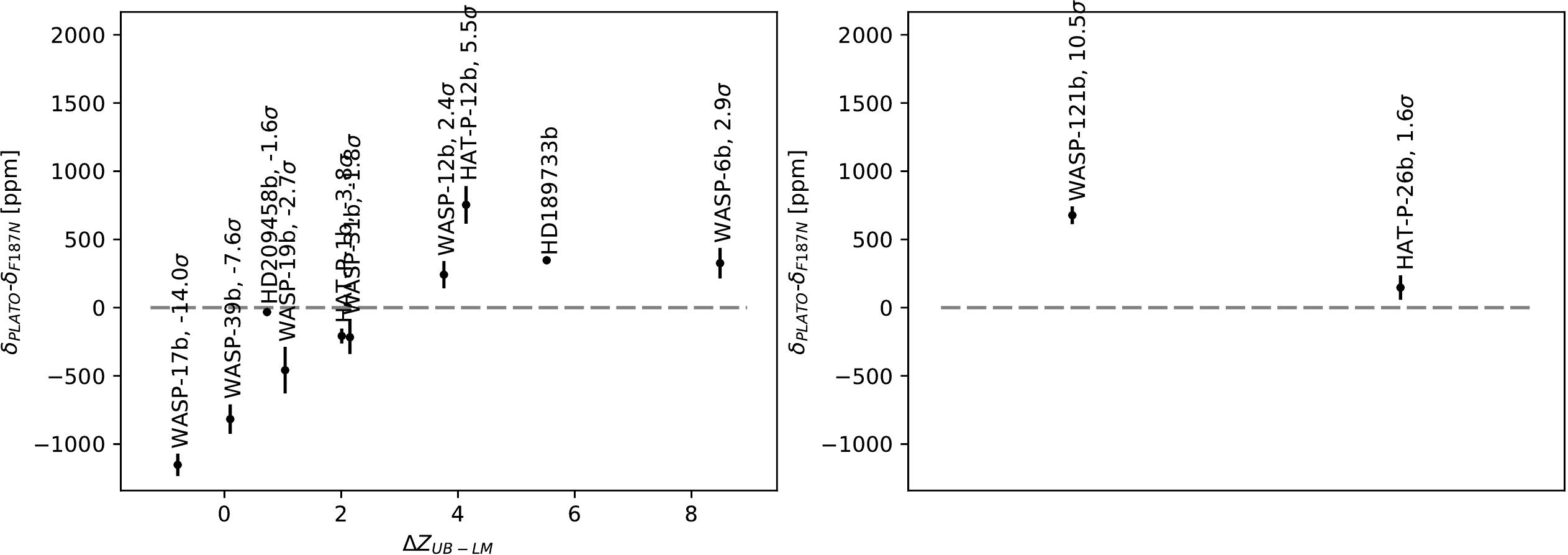}
\caption{Same as in Fig.~\ref{fig:cheops_tess} for the $\delta_{\rm PLATO}-\delta_{\rm F187N}$ transit depth difference. Data points with no error bars are saturated even in the shortest available exposure for NIRCAM.}\label{fig:plato_jw187}
\end{figure*}

\begin{figure*}
\centering
\includegraphics[width=\linewidth]{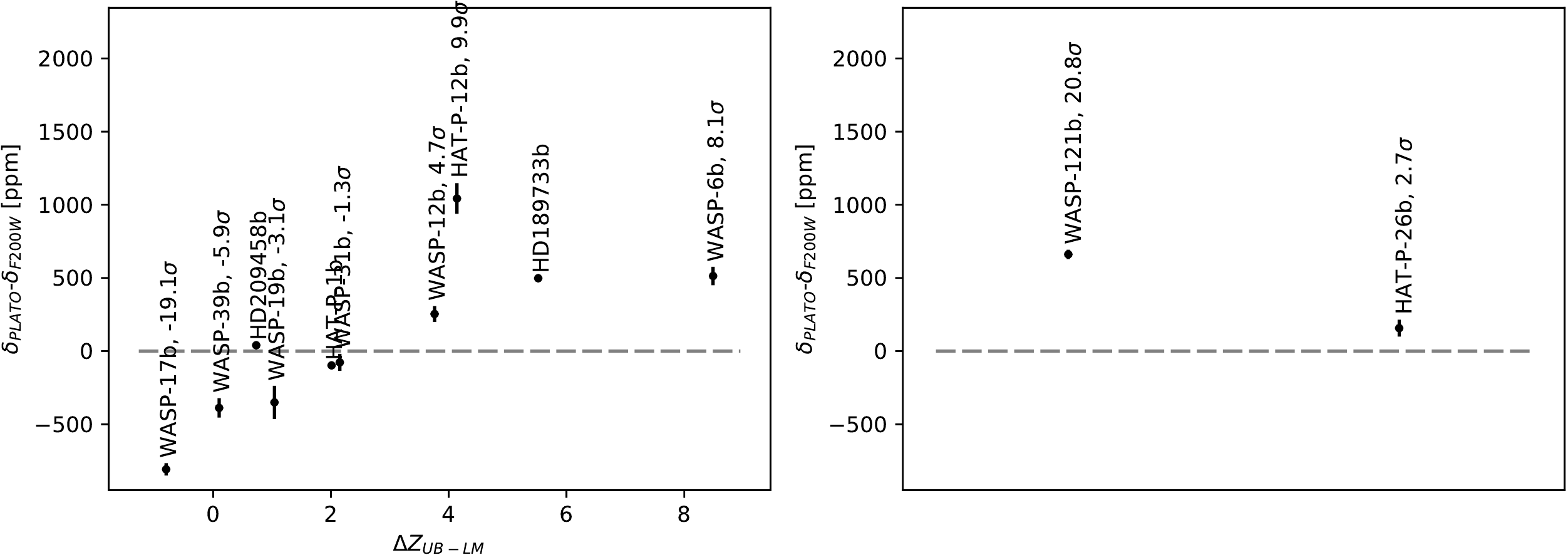}
\caption{Same as in Fig.~\ref{fig:cheops_tess} for the $\delta_{\rm PLATO}-\delta_{\rm F200W}$ transit depth difference. Data points with no error bars are saturated even in the shortest available exposure for NIRCAM.}\label{fig:plato_jw200}
\end{figure*}

\begin{figure*}
\centering
\includegraphics[width=\linewidth]{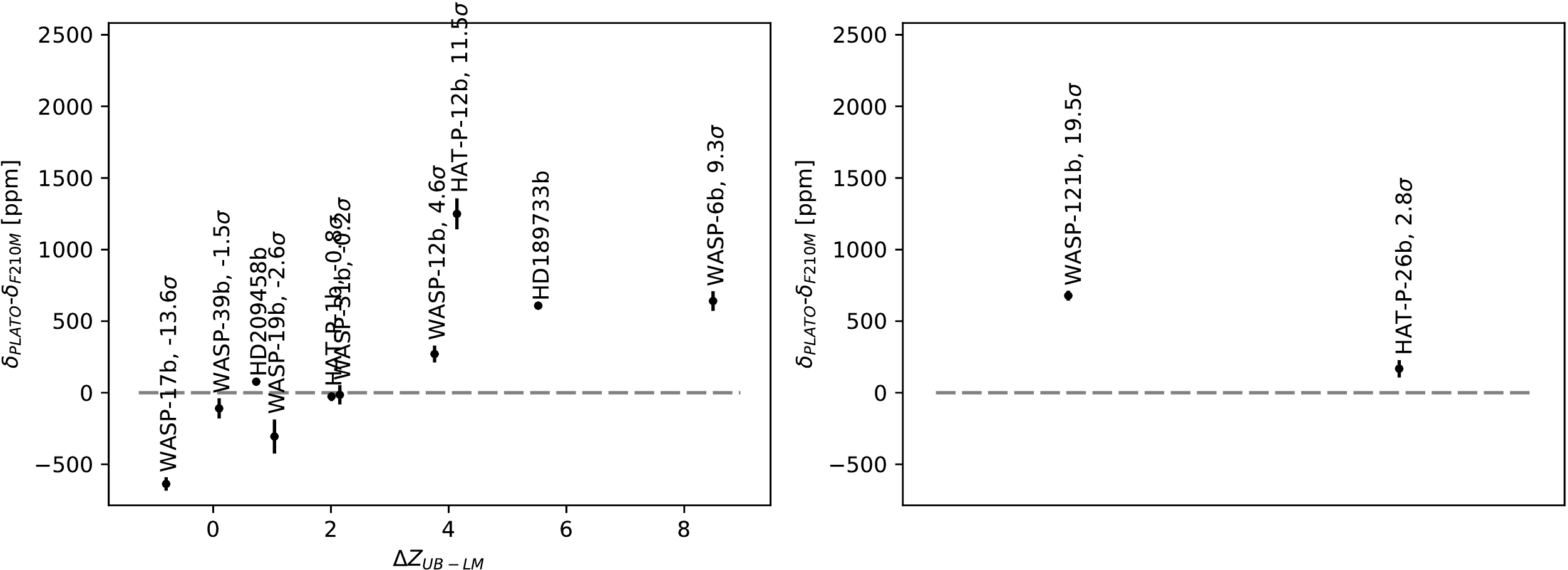}
\caption{Same as in Fig.~\ref{fig:cheops_tess} for the $\delta_{\rm PLATO}-\delta_{\rm F210M}$ transit depth difference. Data points with no error bars are saturated even in the shortest available exposure for NIRCAM.}\label{fig:plato_jw210}
\end{figure*}

\begin{figure*}
\centering
\includegraphics[width=\linewidth]{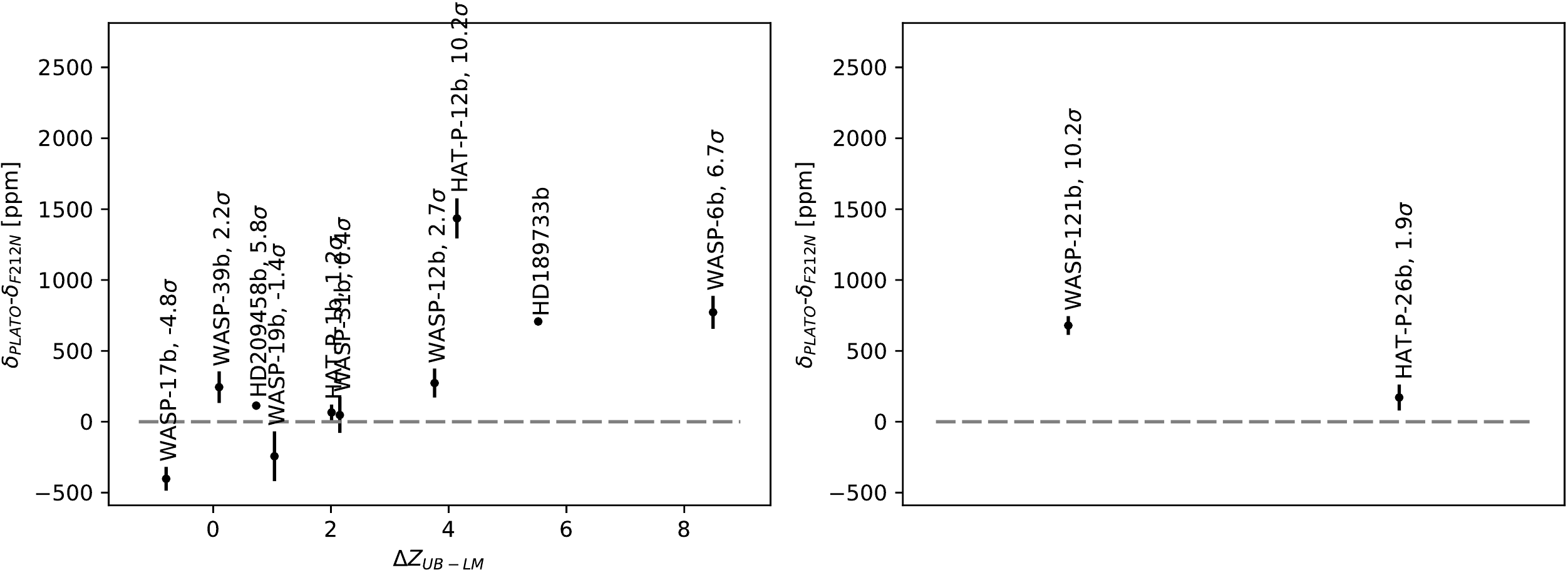}
\caption{Same as in Fig.~\ref{fig:cheops_tess} for the $\delta_{\rm PLATO}-\delta_{\rm F212N}$ transit depth difference. Data points with no error bars are saturated even in the shortest available exposure for NIRCAM.}\label{fig:plato_jw212}
\end{figure*}

Our results are shown in Figs.~\ref{fig:plato_jw070}---\ref{fig:plato_jw212}. We find that for NIRCAM's passbands F070W and F090W the difference in transit depth is lower than 300 ppm. As in the previous case, this is due to the large overlap of these passbands with PLATO's. Nonetheless, the photometric precision of both PLATO and NIRCAM is much better than the ones of CHEOPS and TESS, such that the significance level of the detections is $>3\sigma$ for the most favourable HJs and for the F090W passband. In general we find that the transit depth difference is positive when PLATO is compared with NIRCAM's F090W filter, due to the scattering slope in the optical. Conversely, despite the fact that PLATO and the F070W filter have similar effective wavelengths (Table~\ref{tab:instruments}), the transit depth obtained with the F070W filter is lower than the one returned by PLATO, i.e. the transit depth difference is negative, because PLATO's passband has a long tail towards red wavelengths.


If we shift NIRCAM's observations to longer wavelengths, we find that the transit depth difference can be either positive or negative. A positive difference indicates that the effective radius of the planet as seen by PLATO is larger than the one seen by NIRCAM. This means that the atmosphere of the planet is more opaque at optical wavelengths than in the infrared. Vice versa, a negative transit depth difference indicates that the atmosphere is more opaque at near-IR wavelengths than in the optical. Since water absorption bands are the main source of opacity between 1~\micron\ and 2~\micron, the transit depth difference between PLATO and NIRCAM gives straightforwardly an indication of the strength of the water features in the atmospheres of HJs compared with the opacity in the optical.

In a similar way, \citet{Sing2016} define the index $\Delta Z_{\rm UB-LM}$ which compares the relative strength of scattering, which is strongest at blue-optical wavelengths, to that of molecular absorption mid-infrared wavelengths (mainly H$_2$O, CO and CH$_4$). For this reason the transit depth difference correlates with $\Delta Z_{\rm UB-LM}$ in Figs.~\ref{fig:plato_jw070}---\ref{fig:plato_jw212}. Only in a few cases there is a looser correlation between $\Delta Z_{\rm UB-LM}$ and the transit depth difference, i.e.\ when NIRCAM's filters F115W, F150W2+F162M, F150W2+F164N, F210M and F212N are used. This is due to the fact that these filters encompass spectral regions where the strength of the water bands is lower compared with the other filters (see Fig.~\ref{fig:spectrum}). The filters listed above are thus less effective in the measurement of the water bands absorption.

We remark that the analyzed systems have already been observed in spectrophotometry. By consequence, the approach we outline in this work is redundant with respect to the information already available. Nonetheless, they represent a realistic testbed to predict the output expected from this technique when applied to fainter stars. As a matter of fact, while the signal does not depend on the magnitude of the star $m$, the expected noise increases with increasing $m$. Assuming that photon noise is the dominant source of uncertainty, poissonian statistics leads to the result that the expected noise scales as $10^{m/5}$. Hence, the significance levels reported in Figs.~\ref{fig:cheops_tess}---\ref{fig:plato_jw212} are about 1.6 and 2.5 times smaller if the host stars are 1 and 2 magnitude fainter respectively with respect to the systems that we have analyzed. Our approach is thus still valuable also for stars whose faintness does not allow spectrophotometry.

\section{Effects of stellar activity}\label{sec:activity}
Stellar activity introduces chromatic effects which hamper the measurement of the transit depth \citep{Apai2018}. First of all, if the projection of the HJ against the stellar surface crosses an active region, then the transit light curve shows either a bump or a dip if the active region is spot-dominated or facula-dominated respectively \citep[e.g.\ ][and references therein]{Scandariato2017}. The presence of unocculted active regions is more subtle. For example, the flux contrast of cool starspots against the unperturbed photosphere decreases with increasing wavelength. By consequence, if the projection of the HJ  on the stellar disk does not occult any spot, then the fraction of stellar light blocked by the planet (i.e.\ the transit depth) is larger at shorter wavelengths. Thus, spot-dominated stellar activity may be mistaken as Rayleigh scattering in the atmosphere of the HJ. Conversely, the presence of hot faculae in the photosphere counteracts (and potentially cancels) the decrease of the transit depth with increasing wavelength.

In this section we discuss the effects of stellar activity for different combinations of stellar spectral type, planetary transmission spectra and orbital distance on the measurement of transit depths. In particular, we analyze planetary transits over quiet and active stellar photospheres. To this aim, for each simulated scenario we simulate the out-of-transit and in-transit photometry as done in Sect.~\ref{sec:known}, and we also predict the transit time duration \tduration, which in turns enters in the computation of the uncertainty on the transit depth.

We follow the same approach as in Sect.~\ref{sec:known}, i.e.\ we compute the transit depth measured using couples of  different instrumental setups and looking for significant differences. We do not include WFC3 in this analysis as its exposure time calculator does not allow to simulate in bulk the large number of scenarios we investigate in the current section.


\subsection{Construction of simulated planetary systems}

In our simulations, we consider main sequence stars with \teff\ ranging from 4000~K to 7000~K (corresponding to the FGK spectral types) with a step of 500~K. To model the activity-affected stellar spectra, for each stellar \teff\ we assume the temperature contrast of spots provided by \citet{Berdyugina2005} and extrapolate it up to \teff=7000~K. We also follow \citet{Gondoin2008} and assume that faculae are 100~K hotter than the unperturbed photosphere. For the fractional coverage of the photosphere by spots and faculae,${\rm ff_s}$ and ${\rm ff_f}$ respectively, we assume the filling factors provided by \citet{Rackham2018} as functions of spectral type. Finally, we model the intensity $I(\lambda)$ radiated by the star as:
\begin{equation}
\rm I(\lambda)=(1-ff_s-ff_p)*I_p(\lambda)+ff_s*I_s(\lambda)+ff_f*I_f(\lambda),
\end{equation}
where $I_p(\lambda)$, $I_s(\lambda)$ and $I_f(\lambda)$ are the intensities of the unperturbed photosphere, spots and faculae respectively. As for the unperturbed photosphere (Sect.~\ref{sec:photometry}), we model $I_s(\lambda)$ and $I_f(\lambda)$ using the BT-Settl spectral model with the corresponding temperature.

For each simulated \teff, we also interpolate the temperature--radius--mass-magnitudes relations provided by \citet{Pecaut2013} to obtain the stellar parameters needed for the computation of the photometric accuracy of the observations:
\begin{itemize}
\item the stellar radius R$_{\rm *}$, which enters in the computation of the transit time duration \tduration;
\item the stellar mass M$_{\rm *}$, which sets the orbital velocity and, thus, also \tduration;
\item the stellar absolute magnitudes.
\end{itemize}

From the point of view of the planet, we analyze some specific atmospheres from the CREATES database, based on the expected transit depth difference (see the next sections). For each selected atmospheric model, we naively assume that the planetary equilibrium temperature \teq\ is the only parameter determining the physics of the atmosphere. We thus need to place the simulated HJs at the correct distance from the simulated host star to keep \teq\ fixed. To this purpose, we solve Eq.~6.60 in \citet{Perryman2011} for the semimajor axis $a$ and we obtain:
\begin{equation}
\rm a=\frac{R_*}{2}\left(\frac{T_{eff}}{T_{eq}}\right)^2\left[f\cdot(1-A)\right]^\frac{1}{2},\label{eq:a}
\end{equation}
where $A$ is the Bond albedo of the planet, $f$ describes the effectiveness of circulation, \teff\ is the stellar effective temperature and $a$ is the semi-major axis corresponding to \teq. 

Finally, to write \tduration\ in terms of stellar, planetary and orbital parameters, we combine Eqs.~6.16 and 6.28 in \citet{Perryman2011} and obtain that the transit duration is given by:
\begin{equation}
\rm \frac{t_d}{t_d^\prime}=\left(\frac{M_*^\prime}{M_*}\frac{a}{a^\prime}\right)^\frac{1}{2}\frac{R_*+R_p}{R_*^\prime+R_p},\label{eq:transitTime}
\end{equation}
where the primed quantities refer to the parameters of the original systems to which the selected HJs belongs to. If we plug Eq.~\ref{eq:a} into Eq.~\ref{eq:transitTime}, we finally obtain:
\begin{equation}
\rm \frac{t_d}{t_d^\prime}=\left(\frac{M_*^\prime}{M_*}\frac{R_*}{R_*^\prime}\right)^\frac{1}{2}\left(\frac{R_*+R_p}{R_*^\prime+R_p}\right)\frac{T_{eff}}{T_{eff}^\prime},
\end{equation}
which allows to scale the transit duration of the original planetary system to the simulated scenario. For simplicity, we take $\rm R_p$ as the median $\rm R_p(\lambda)$ between 0.4~\micron\ and 0.6~\micron.

We thus end up with a collection of simulated planetary systems, each one with its own set of stellar (\teff, $\rm ff_s$, $\rm ff_p$, V-band absolute magnitude) and planetary (\tduration, transmission spectrum) parameters needed to compute the expected transit depths and the corresponding uncertainties, as described in Sect.~\ref{sec:known}.

\subsection{CHEOPS-TESS comparison}

\begin{figure}
\centering
\includegraphics[width=\linewidth]{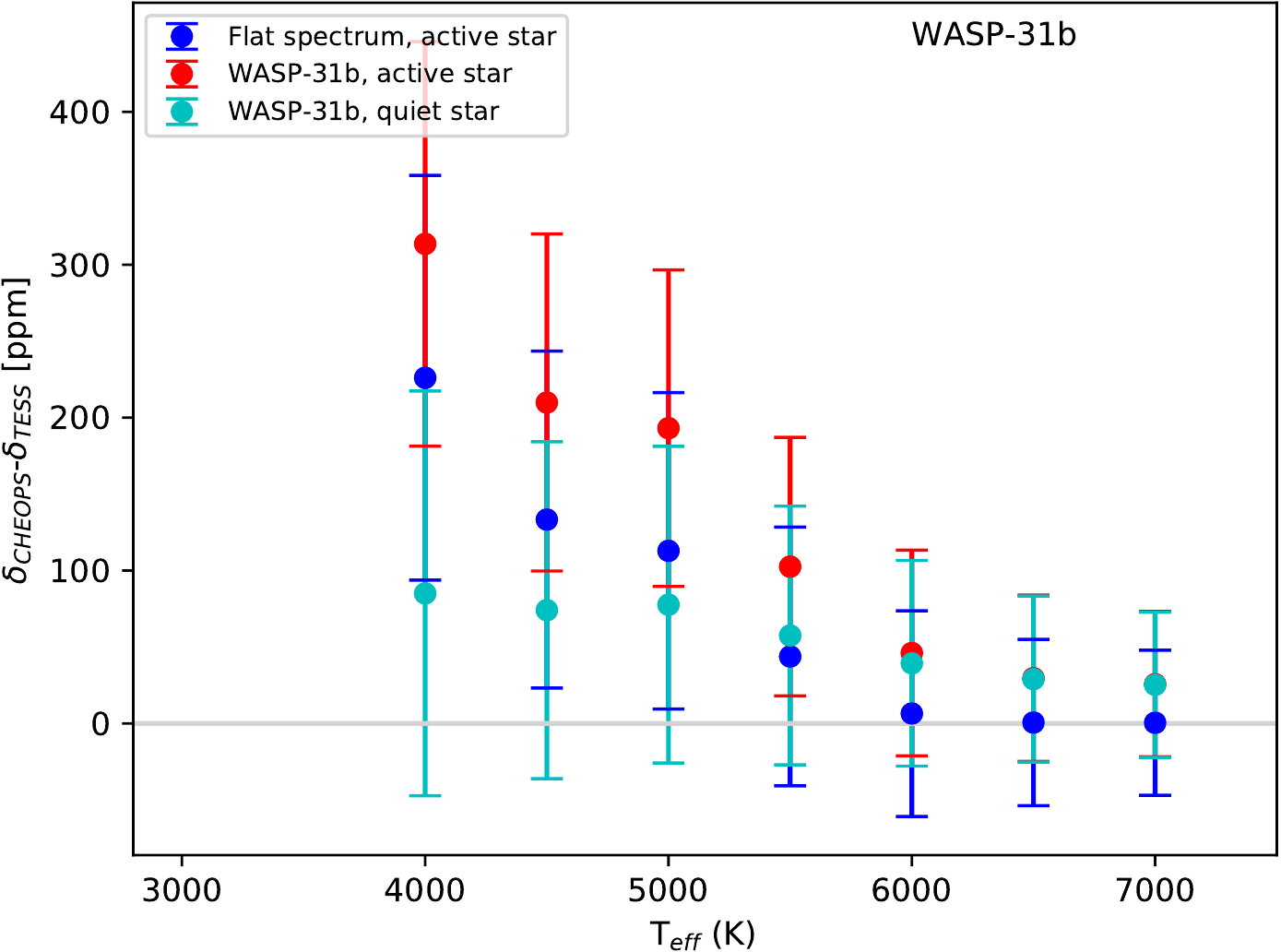}
\includegraphics[width=\linewidth]{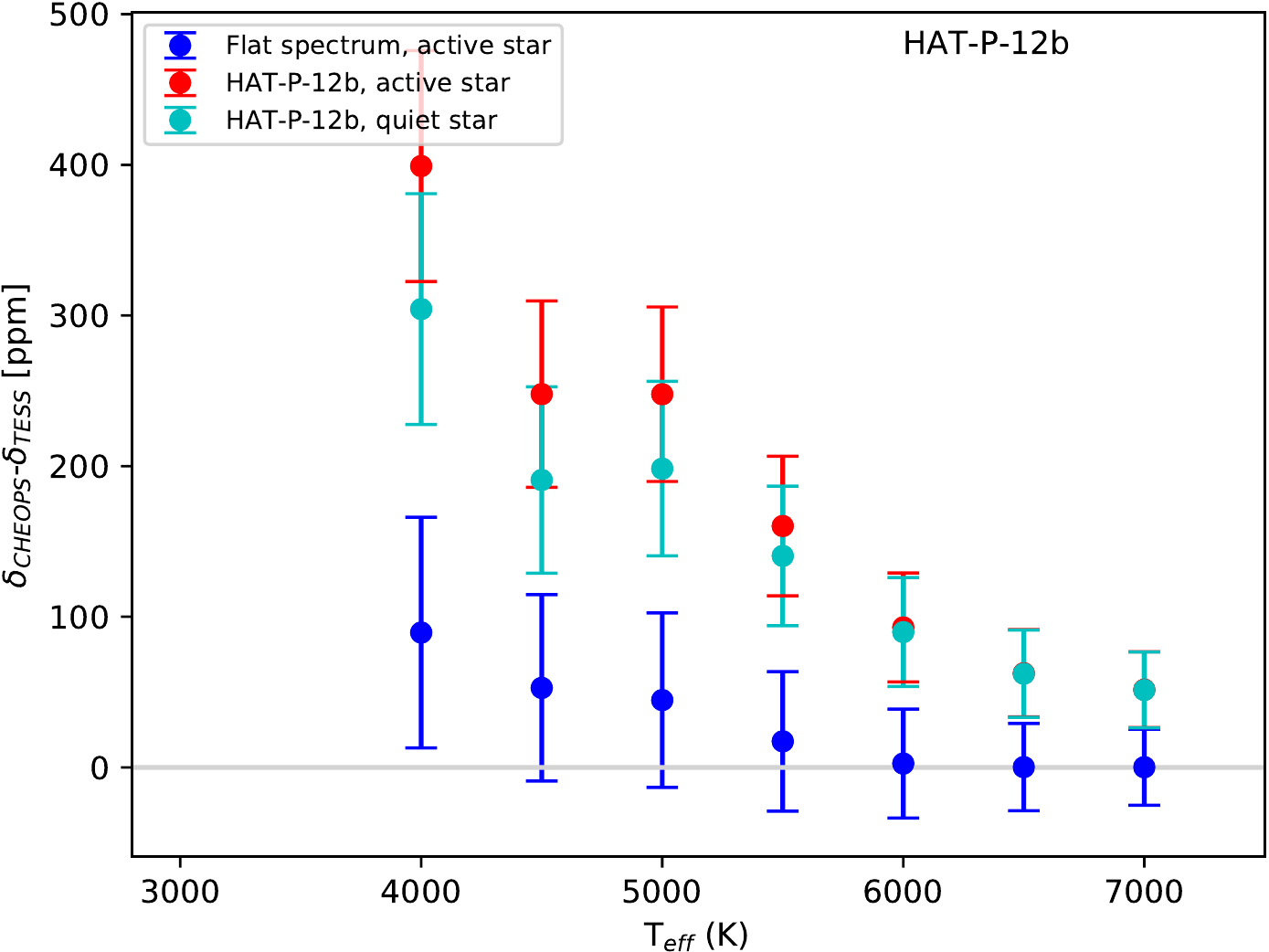}
\caption{Simulation of the $\delta_{CHEOPS}-\delta_{TESS}$ for host stars with V=8, \teff\ between 4000~K and 7000~K and atmospheric models as WASP-31b (top panel) and HAT-P-12b (bottom panel). In each panel, cyan symbols represent the simulation assuming no stellar activity, while red symbols include the effect of stellar activity. Blue symbols represent the simulation of a HJ with a flat transmission spectrum transiting an active star.}\label{fig:cheops_tess_sim}
\end{figure}

In Sect.~\ref{sec:known} we found that the HJs in the CREATES database are not suited for a comparative analysis using CHEOPS and TESS, mainly because the corresponding host stars are too faint to get high precision photometry with the two instruments. We now simulate the same HJs, rescaling the planetary system as discussed above, depending on the \teff\ of the host star. In particular, we focus on the atmospheric models of WASP-31b and HAT-P-12b, as they show the minimum and maximum transit depth difference $\delta_{CHEOPS}-\delta_{TESS}$ respectively (Fig.~\ref{fig:cheops_tess}). This choice aims at analyzing the effects of stellar activity in the range of the planetary atmospheres  in Table~\ref{tab:hjs}. We also set the apparent magnitude of the simulated stars to V=8, which is within the saturation limits of both CHEOPS and TESS. For each star-HJ scenario, we compute the transit depths and the corresponding uncertainties as in Sect.~\ref{sec:photometry}. The results of our simulations are shown in Fig.~\ref{fig:cheops_tess_sim}.

For the case of WASP-31-b, we obtain that the $\delta_{CHEOPS}-\delta_{TESS}$ difference is consistent with the null detection within 1$\sigma$, for both scenarios with a quiet and active host star (cyan and red symbols respectively in Fig.~\ref{fig:cheops_tess_sim}). If anything, we find that the expected signal increases with decreasing stellar \teff, especially in the \lq\lq active star\rq\rq\ scenario. To verify if this is due to stellar activity, we simulate the same system substituting the planetary transmission spectrum with a flat spectrum, matching the transit depth of the flat and model spectra in the visible band. We find that the stellar activity signal is $\lesssim$200~ppm, negligible at the 1$\sigma$ level (blue symbols). We also find that, in the presence of a scattering atmosphere, the atmospheric signal is enhanced, still within the error bars.

For HAT-P-12b we find a larger atmospheric signal, as expected, with a significance level of a few $\sigma$, especially for \teff$<$6000~K. We also find that the stellar activity signal is consistent with the null detection within 1$\sigma$. Finally, also in this case we find that stellar activity pushes up the atmospheric signal, introducing a bias smaller than the measurement uncertainties.

\subsection{PLATO-NIRCAM comparison}

To compare the transit depths as observed by PLATO and NIRCAM@JWST, we analyze WASP-17b and HAT-P-12b, whose transit depth difference between PLATO and NIRCAM's passbands are the lowest and the highest respectively (Figs.~\ref{fig:plato_jw070}---\ref{fig:plato_jw212}). To avoid saturation of the instruments, we simulated host stars with V=13.

\begin{figure}
\centering
\includegraphics[width=\linewidth]{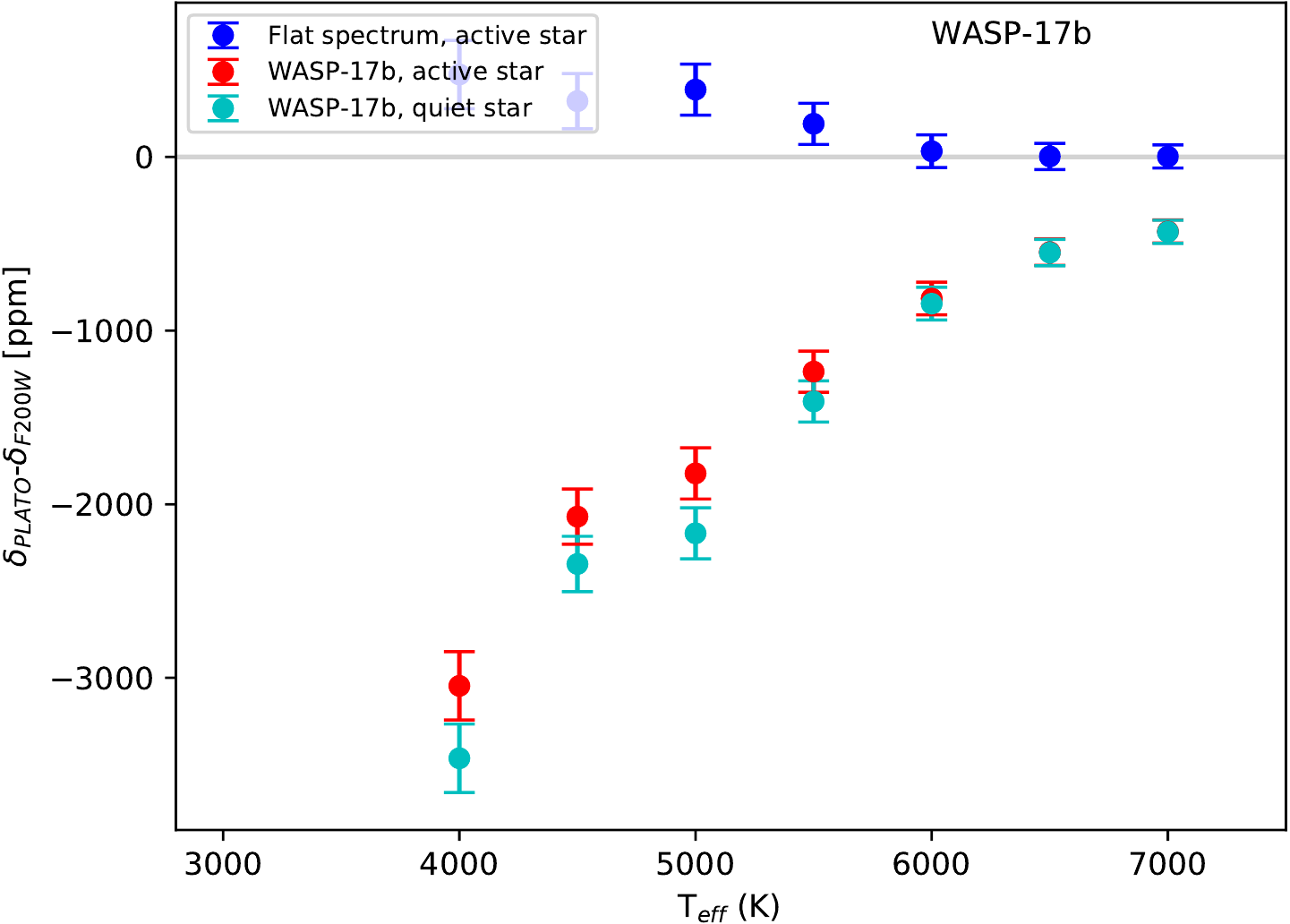}
\includegraphics[width=\linewidth]{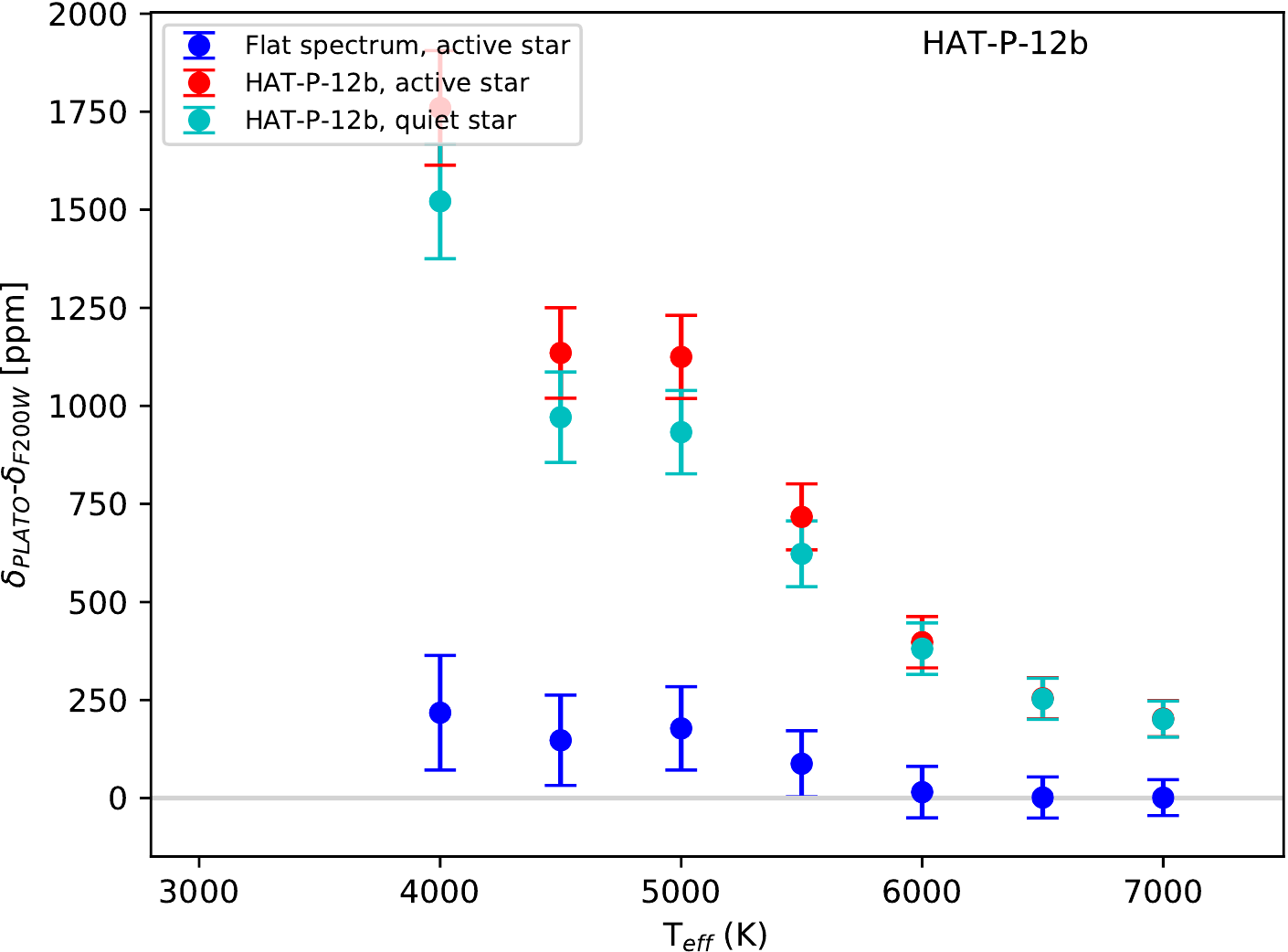}
\caption{Same as in Fig.~\ref{fig:cheops_tess_sim}, for a WASP-18b-like (top panel) and a HAT-P-12b-like (bottom panel) HJ orbiting stars with V=13 and different \teff.}\label{fig:plato_jw_sim}
\end{figure}

For both planets, we find that the transit depth measurements and its dependence on spectral type changes with the NIRCAM filter used for the observations. In most cases, the significance of the atmospheric signal is high, ranging from $\sim5\sigma$ for \teff=7000~K to a few tens of $\sigma$ for cooler \teff. We also find that the significance increases with the $\rm\lambda_{eff}$ of the selected NIRCAM passband. The only exception to this general rule is $\delta_{\rm PLATO}-\delta_{\rm F070W}$, because the wavelength leverage between PLATO and the filter F070W is too short to detect any wavelength-dependence significantly. In Fig.~\ref{fig:plato_jw_sim} we show the case of $\delta_{\rm PLATO}-\delta_{\rm F200W}$ for the two discussed model atmospheres.

Our simulations also show that, for both model atmospheres, stellar activity is expected to produce a negligible signal in the case of a flat transmission spectrum, as we already found analyzing the CHEOPS-TESS comparison. Moreover, the effect of stellar activity is to push up the atmospheric signal, but again this bias is within the measurement uncertainties, thus negligible.

The main difference between the two models is the sign of the atmospheric signal. For HAT-P-12b the difference is positive, which confirms the scattering-dominated transmission spectrum. Conversely, the spectrum of WASP-17b is characterized by strong water absorption bands at $\sim$1~\micron, $\sim$1.5~\micron\ and $\sim$2~\micron, where most of NIRCAM's passbands are centered (Fig.~\ref{fig:spectrum}). This leads to the fact that that the atmosphere is more opaque at near-infrared wavelengths than in the optical, thus leading to negative transit depth differences. The only exceptions are the F150W2+F162M and F150W2+F164N, which are centered between the two water absorption bands at $\sim$1.5~\micron and $\sim$2~\micron. In these cases, the transit depth measured by PLATO and NIRCAM are comparable, such that no difference can be detected with enough significance.

Between the two extreme cases represented by WASP-17b and HAT-P-12b there are other atmospheric models which lead to weaker signals. For example, for the atmospheric model of HAT-P-1b, whose transit depth difference is close to zero in Figs.~\ref{fig:plato_jw070}---\ref{fig:plato_jw212}, we find that the signal is too weak for a clear detection regardless of the spectral type of the host stars. Moreover, the stellar activity signal is of the same order of magnitude of the expected atmospheric signal, which means that it is hard to discriminate whether a detection is of stellar or planetary origin, provided the required photometric precision.

\section{Discussion}
In this paper we have outlined a strategy to measure the effects of scattering/absorbing planetary atmospheres onto the transit depths of Hot Jupiters (HJs) using current and future space telescopes: HST, TESS, CHEOPS, PLATO and JWST. This technique consists in measuring and comparing the transit depth measured in different photometric bands, as done in, e.g., \citet{Nascimbeni2013}, \citet{Mallonn2015} and \citet{Nascimbeni2015}. This work is thus a feasibility study for the application of the photometric multi-band analysis with space-borne facilities.

First of all we analyzed the set of known HJs in the CREATES database, simulating the observations done with online and forthcoming facilities such as WFC3@HST, TESS, CHEOPS, PLATO and NIRCAM@JWST. We obtained that the CHEOPS-TESS synergy cannot lead to significant results because their respective passbands overlap with each other. This means that the wavelength leverage between the two telescopes is too small to measure the spectral slope of the atmosphere. Moreover, the host stars are too faint for the two telescopes to obtain photometric light curves with the required precision. The expected signal may become statistically significant if the observations of either CHEOPS or TESS are compared with the ones of WFC3 used in the UV channel.

Conversely, the PLATO-NIRCAM synergy is more promising, both because the two instruments do not overlap in wavelength (except for NIRCAM's F070W and F090W) and because of their larger sensitivity. Our results show that the atmospheric signal can be measured with strong significance across the range of analyzed HJs. The only exceptions are those HJs whose atmospheres have comparable optical depths in PLATO's and NIRCAM's passbands due to matching opacities in the optical and water molecular absorption bands.

In the perspective of new HJ discoveries provided by TESS and PLATO, we also simulated different scenarios where host stars of different \teff\ and activity levels are orbited by HJs with the atmospheres in the CREATES database. We find that the CHEOPS-TESS and PLATO-NIRCAM synergies can effectively provide information on the atmospheres of HJs with strong significance, provided that the host star is bright enough to achieve the required photometric precision. In particular, we find that the atmospheric effect is maximized at later spectral types, and that the typical stellar activity level of main sequence stars negligibly contributes to the difference of transit depths.

We remark that our work is based on the analysis of mono-transits. If multiple transits are observed, than uncertainties on the transit depths measurements are expected to decrease with the square root of the number of observed transit. If stellar brightness is an issue in terms of photometric precision, then in principle it is possible to observe more transits and achieve better significance on the detection. Nonetheless, in this case a robust model is needed to correct the stellar variability, which significantly biases the multi-epoch observations of transits.

On the bright side of the stellar population in the solar neighborhood, we notice that JWST large aperture makes stars brighter than K$_{\rm S}\simeq$8-9 saturate even in the shortest exposure time available \citep{Stansberry2014}. This problem can be mitigated using NIRCAM's weak lens WL+8, which defocuses the star spreading the light out over many pixels. Thus, while this instrumental setup extends the applicability of our approach to stars as bright as K$_{\rm S}\simeq$5 \citep{Stansberry2014}, it also decreases the uncertainties due to the telescope's jitter. The latest version of the ETC (v. 1.3) does not allow the simulation of this instrumental setup. When it becomes available, it will be possible to redo the calculations in this work, and in particular the estimate of the maximum integration time before saturation and the corresponding photometric noise.

We have not included ARIEL in this preparatory study as its development is still in a preliminary stage and technical details such as the passband still needs to be confirmed. Nonetheless, we remark that it will be able to provide accurate photometry in the visual, red and near-infrared bands \citep{Tinetti2018}. Hence, ARIEL will be an important candidate for the synergetic observations of HJs' primary transits outlined in this paper, as its spectral coverage nicely overlaps with the ones analyzed in this work.

Finally, we remark that our aim in this work is to predict the performances of the multi-band photometry approach in the atmospheric characterization of HJs. To this purpose, we do not investigate the whole space of atmospheric parameters, which may lead to unrealistic atmospheric models. We rather choose to analyze the model atmospheres which fit the spectrophotometric data collected so far for a dozen of HJs, taking them as a good representative sample of the family of hot giant planets. In this sense, our analysis cover a restricted though realistic range of the atmospheric parameter space.

\section*{Acknowledgements}

G.S. acknowledges financial support from \lq\lq Accordo ASI--INAF \rq\rq\ No. 2013-016-R.0 July 9, 2013 and July 9, 2015.






\bibliographystyle{mnras}
\bibliography{References}


%
%
%
%

\bsp	
\label{lastpage}
\end{document}